# Enhanced Spin Lifetime and Long-Range Spin Transport in p-Silicon using Spin Gapless Semiconductor as Ferromagnetic Injector

**Nilay Maji\*, Subham Mohanty, Pujarani Dehuri, Garima Yadav**

Department of Physics and Astronomy, National Institute of Technology Rourkela, Rourkela-769008, Odisha, India

\*Author to whom correspondence should be addressed: majin@nitrkl.ac.in

## Abstract

Electrical spin injection and transport in silicon are central challenges for realizing semiconductor-based spintronic devices, particularly in p-type Si, where strong spin relaxation and interface effects often suppress detectable spin signals. Here, we report electrical spin injection, accumulation, and transport in lightly doped p-type silicon using the spin-gapless Heusler compound $Mn_2CoAl$ as a ferromagnetic spin injector, separated from the p-Si channel by a thin MgO tunnel barrier in a lateral device geometry. Spin transport is systematically investigated through three-terminal (3-T) Hanle and four-terminal (4-T) nonlocal (NL) spin-valve and Hanle measurements. Clear Lorentzian Hanle signals are observed in the 3-T configuration from 5 K up to room temperature, yielding a spin lifetime of ~0.68 ns at 300 K that increases to ~4.11 ns at 5 K. Temperature-dependent analysis reveals a weak power-law dependence of the spin lifetime, indicating Bir–Aronov–Pikus–type spin relaxation mechanism. To validate genuine spin transport, NL spin-valve and Hanle measurements were performed, revealing well-defined spin-valve switching and controlled spin precession at 5 K. From NL Hanle fitting, a spin lifetime of ~5.65 ns and a spin diffusion length of ~0.82 μm are extracted, confirming diffusive long-range spin transport in the p-Si channel. Although NL signals diminish at elevated temperatures due to reduced interfacial spin polarization and thermal noise, the combined 3-T and 4-T results establish spin-gapless $Mn_2CoAl$ as an effective spin injector for p-type silicon. These findings highlight the potential of spin-gapless semiconductors for improving spin injection efficiency and advancing Si-compatible spintronic devices.

Electrical spin injection into semiconductors[1-5] has become a central issue in modern spintronics due to its direct relevance to low-power logic devices,[6,7] nonvolatile memories,[8,9] and hybrid CMOS[10,11] spintronic architectures. Integrating spin functionality with established semiconductor platforms is widely regarded as a key pathway toward energy-efficient information processing and logic-in-memory concepts. In this context, achieving reliable electrical spin injection, transport, and detection in technologically mature semiconductors such as silicon, particularly in p-type Si, remains a timely and actively pursued challenge. Silicon is particularly attractive because of its compatibility with existing microelectronics and its long intrinsic spin diffusion length.[12,13] However, electrical spin injection into Si is hindered by conductivity mismatch[14,15] and significant spin depolarization[16,17] at ferromagnet/semiconductor interfaces. Early experimental studies predominantly utilized three-terminal (3-T) Hanle measurements,[18,19] which established evidence of spin accumulation largely confined to n-type silicon. It is now well recognized, however, that 3-T signals can be affected by spurious contributions such as anisotropic magnetoresistance,[20,21] local Hall effects, and interface-related phenomena, complicating the interpretation of true spin transport. The four-terminal (4-T) nonlocal geometry[3,22] has therefore been established as the most stringent method for demonstrating genuine spin transport in semiconductors, as it spatially separates charge and spin currents and minimizes parasitic magnetoresistive effects. Nonlocal Hanle measurements[23,24] are widely accepted as a benchmark for validating spin injection and transport. Nevertheless, achieving robust nonlocal spin signals, particularly in p-type silicon, remains experimentally challenging due to strong hole-mediated spin relaxation and enhanced interfacial spin depolarization at ferromagnet/Si interfaces, highlighting the need for improved spin injector materials and interfaces.

From a materials perspective, conventional ferromagnetic metals and half metallic alloys have been extensively explored for spin injection into semiconductors. While these materials offer relatively high spin polarization, their performance is often limited by interfacial disorder,[25,26] Fermi-level mismatch,[27] and thermally induced depolarization[28] at the semiconductor interface. These limitations have motivated interest in alternative spin sources that can provide highly spin-polarized carriers while maintaining better compatibility with semiconducting channels. Spin-gapless semiconductors[22,29-33] have recently emerged as a promising class of spin injectors, characterized by a zero bandgap in one spin channel and a finite gap in the opposite channel. This unique electronic structure enables, in principle, highly spin-polarized carrier injection with semiconducting carrier densities, offering potential advantages over both conventional ferromagnets and half-metals. $Mn_2CoAl$ (MCA)[29,34-36] is a prototypical spin-gapless Heusler compound that has attracted considerable attention owing to its high spin polarization, low carrier concentration, and favorable band alignment with semiconductors, making it an appealing candidate for silicon-based spintronic devices. In this work, we investigate electrical spin injection from spin-gapless $Mn_2CoAl$ into a p-type silicon channel via thin MgO tunnel barrier using lateral device geometry. The structural, magnetic, and transport characteristics of MCA film have already been reported in our another work.[37] Here, spin transport is examined through a direct comparison of 3-T Hanle and 4-T nonlocal Hanle measurements. Clear lorentzian Hanle signals observed in both geometries provide unambiguous evidence of spin transport in p-Si driven by a spin-gapless injector.

Lateral spin transport devices were fabricated on silicon-on-insulator substrates with a p-type Si channel having doping density of $4.5\times10^{16}$ cm$^{-3}$. Prior to deposition, the native oxide on the Si surface was removed using diluted HF, followed by rinsing in deionized water and isopropyl alcohol. Spin injector and detector contacts were formed by depositing $Mn_2CoAl$/MgO bilayers using magnetron sputtering. The MgO tunnel barrier (~2 nm) was deposited by rf sputtering, while the $Mn_2CoAl$ layer was deposited by dc sputtering, without breaking vacuum. Device patterning was carried out using electron-beam lithography followed by mesa etching. Two ferromagnetic tunnel contacts (labeled "2" and "3") were defined by a lift-off process, with a lateral size of approximately $10 \times 20$ μm$^2$ and $20 \times 30$ μm$^2$. Finally, the pad electrodes for contacts "1"-"4", Au (100nm)/Cr (50nm), were fabricated by lift-off method. The separations between contacts "1"-"2" and "3"-"4" were 80 μm, much larger than the expected spin diffusion length in p-Si, while the injector–detector separation (contacts "2"-"3") in the nonlocal geometry was ~1 μm. Spin transport measurements were performed using both three-terminal (3-T) and four-terminal (4-T) nonlocal configurations, as shown in Fig. 1(a). A spin-polarized current was injected from contact "2" to contact "1". In the nonlocal geometry, the spin signal was detected as a voltage between contacts "3" and "4", whereas in the 3-T configuration, the spin accumulation voltage was measured between contacts "2" and "4". An in-plane magnetic field along the ±y direction was applied to switch the magnetization between parallel and antiparallel states, while Hanle spin precession measurements were carried out by applying an out-of-plane magnetic field along the ±z direction.

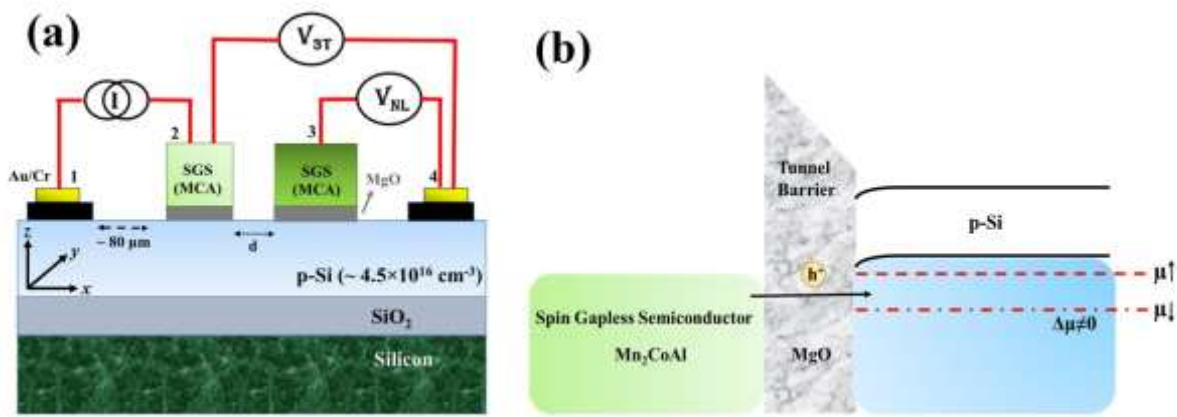

**Figure 1. (a)** Schematic diagram of the Hanle (3-T and NL) measurement setup. **(b)** The schematic energy band profile of the heterojunction and spin accumulation in the valence band of p-Si.

At first, electrical spin injection from the ferromagnetic spin gapless semiconductor Mn$_2$CoAl (MCA) thin film into p-type silicon was investigated using 3-T Hanle geometry. A constant dc current (200 µA) was applied between terminals "2" and "1", as shown schematically in Fig. 1(a). This current injection creates a nonequilibrium imbalance in the hole population in the valence band of p-Si (Fig. 1(b)), which can be expressed as

$$\Delta\mu = \mu\uparrow - \mu\downarrow \qquad (1)$$

where, µ↑ and µ↓ are the electrochemical potentials of up-spin and down-spin holes, respectively. The in-plane magnetization direction of the Mn$_2$CoAl film determines the orientation of the spin-polarized holes accumulated in the p-Si channel beneath the tunnel contact. When we apply an external magnetic field B perpendicular to the accumulated spin orientation in the semiconductor, the spins undergo Larmor precession[38] with angular frequency

$$\omega = g\mu_B B/\hbar \qquad (2)$$

The precessional dephasing of spins reduces the net spin accumulation, resulting in a characteristic Lorentzian decay of the spin signal.

This spin accumulation detected by three-terminal Hanle voltage (between contacts "2" and "4") can be expressed, as

$$\Delta V_{3T} = V(B) - V(B=0) = \frac{\gamma \Delta\mu}{2e} \qquad (3)$$

where, γ is the tunnel spin polarization of the Mn$_2$CoAl/SiO$_2$ interface.[38] Unlike four-terminal nonlocal geometries, the three-terminal configuration probes the spin accumulation directly underneath the ferromagnetic tunnel contact, where the spin density is maximum.

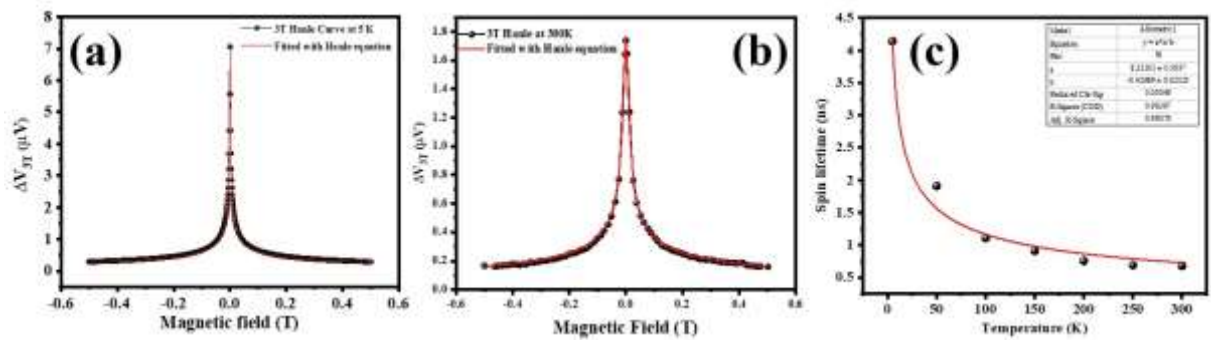

**Figure 2. (a)-(b)** Hanle signal recorded at 5 K and 300 K in conventional 3-T Hanle device. The solid curves in (a) and (b) correspond to the best fit to Eq. (4). **(c)** Temperature dependence of spin lifetime recorded in 3-T Hanle geometry.

Clear Hanle voltage signals with a nearly Lorentzian line shape were observed over a wide temperature range, confirming the simultaneous realization of spin injection, controlled precessional dephasing, and electrical detection. Figure 2(a) and (b) show a representative Hanle curve measured at 5 K and at room temperature. As the magnitude of the transverse magnetic field increases, the precessional dephasing progressively suppresses the spin accumulation, reducing $\Delta V_{3T}$ toward zero. The experimental Hanle curves were fitted using Eq. (4) as written below

$$\frac{\Delta V(B)}{I} = \pm \frac{P^2 \lambda_N}{2\sigma S} exp\left(-\frac{d}{\lambda_N}\right) (1 + \omega^2 \tau_s^2)^{-\frac{1}{4}} \times exp\left\{-\frac{d}{\lambda_N}\left[\sqrt{\frac{1}{2}\left(\sqrt{1 + \omega^2 \tau_s^2} + 1\right)} - 1\right]\right\} cos\left\{\frac{\tan^{-1}(\omega \tau_s)}{2} + \frac{d}{\lambda_N}\left[\sqrt{\frac{1}{2}\left(\sqrt{1 + \omega^2 \tau_s^2} - 1\right)}\right]\right\} \quad (4)$$

Here, P is the spin polarization, $\lambda_N$ is the spin diffusion length, S is the cross sectional area of the detector contact, d is the distance between injector and detector, $\sigma$ is the conductivity of the semiconductor channel, $\tau_s$ is the spin lifetime of the accumulated carriers in the semiconductor channel. During fitting, we set the parameter d = 0 for 3-T case. At room temperature, the extracted spin lifetime in our lightly doped p-type silicon reaches ~0.68 ns, increasing to 4.11 ns at 5 K, which is substantially larger than most values previously reported for silicon using conventional ferromagnetic tunnel contacts. Earlier three-terminal Hanle studies employing metallic CoFe injectors have typically yielded room-temperature spin lifetimes on the order of ~0.16 ns.[39] The pronounced enhancement observed in the present work underscores the effectiveness of the spin-gapless $Mn_2CoAl$ Heusler alloy in enabling efficient spin injection while minimizing extrinsic spin dephasing effects. Notably, spin lifetimes of similar magnitude (~0.16 ns) have also been reported in p-type silicon with comparable doping levels using ferromagnetic-resonance-driven spin pumping,[40] indicating consistency across different experimental approaches.

To further elucidate the dominant spin relaxation mechanism in p-type silicon, temperature-dependent three-terminal Hanle measurements were performed over a wide temperature range. The spin lifetime decreases monotonically with increasing temperature and is well described by a power-law dependence $\tau_s(T)=a.T^b$, yielding an exponent $b = -0.43$, close to $-0.5$ within experimental uncertainty as displayed in Fig. 2(c). This behavior deviates significantly from the Elliott–Yafet mechanism[41,42] expected for bulk silicon, which predicts a much stronger temperature dependence ($\tau_s \propto T^{-2.5}$, where p is the hole concentration) under phonon-limited scattering. The observed $T^{-0.5}$ scaling is instead consistent with Bir–Aronov–Pikus–type spin relaxation[43,44] in p-type silicon, arising from exchange interaction between electron spins and valence-band holes, for which the spin relaxation rate scales as $1/\tau_s \propto pT^{1/2}$. In addition, interface-related spin-flip scattering at the $Mn_2CoAl/SiO_2$/p-Si junction can contribute a similar $T^{1/2}$ dependence, as spins detected in the three-terminal geometry are localized beneath the ferromagnetic tunnel contact and are therefore sensitive to interfacial disorder and fluctuating local fields. The deviation from Elliott-Yafet scaling thus indicates that spin relaxation in the present p-Si devices is governed predominantly by hole-mediated and interface-dominated mechanisms rather than bulk phonon-limited processes, while still allowing efficient room-temperature spin injection using the spin-gapless $Mn_2CoAl$ injector.

However, it is important to note that three-terminal Hanle measurements are inherently sensitive to local and spatially inhomogeneous magnetostatic fringe fields arising from the ferromagnetic tunnel contact. Such fringe fields can artificially broaden the Hanle line shape, leading to an underestimation of the true spin lifetime. As a result, spin lifetimes extracted from the three-terminal geometry should be regarded as a lower bound of the intrinsic spin lifetime in the semiconductor channel. Determination of the upper bound, or the intrinsic transport spin lifetime, requires four-terminal nonlocal Hanle measurements, which probe spin diffusion away from the injection contact and are largely immune to local magnetostatic effects.

Motivated by this consideration, we subsequently investigated spin transport using the four-terminal nonlocal Hanle geometry to validate genuine spin injection into the p-Si channel. Clear nonlocal Hanle signals were observed at low temperature, confirming spin transport in the semiconductor. However, at elevated temperatures above approximately 50 K, the nonlocal signals were strongly suppressed and dominated by noise. This behavior is attributed to thermal degradation of the spin polarization at the $Mn_2CoAl/SiO_2$/p-Si interface and the inherently small magnitude of nonlocal spin signals, which become increasingly susceptible to thermal fluctuations and electrical noise at higher temperatures. Similar limitations of nonlocal spin detection at elevated temperatures have been reported in silicon-based spin transport studies and underscore the critical role of interface quality and thermal stability in achieving room-temperature nonlocal spin transport.

The NL measurements were carried out using a standard ac lock-in technique with a lock-in frequency of 333 Hz, while applying an in-plane magnetic field along the y direction. For NL spin valve measurements, the magnetic field was applied along the y direction, whereas for NL-Hanle measurements, the field was applied perpendicular to the device plane (along the z direction). Prior to measurement, the magnetization directions of the ferromagnetic electrodes at contacts "2" and "3" were initialized by applying an in-plane magnetic field along the y direction. A parallel magnetization configuration was obtained by applying a field of approximately 1000 Oe, while an antiparallel configuration was achieved by reducing the field to about 200 Oe after an initial field of 1000 Oe had been applied.

Depending on the relative magnetization orientation of the injector and detector electrodes, contact 3 probes the electrochemical potential associated with either the majority or minority spin population, reflecting the spin accumulation in the p-Si channel. Clear nonlocal spin-valve signals are observed at magnetic fields of approximately +236 Oe (-205 Oe) and +517 Oe (-527 Oe), corresponding to the coercive fields of the two $Mn_2CoAl$ ferromagnetic electrodes as shown in Fig. 3(a). Although both electrodes are composed of the same $Mn_2CoAl$ material, their different geometry (intentionally introduced during fabrication) leads to distinct switching fields, enabling the realization of a spin-valve effect using identical ferromagnetic materials. The observation of this well-defined nonlocal spin-valve signal provides direct evidence of electrical spin injection, transport, and detection in the silicon channel.

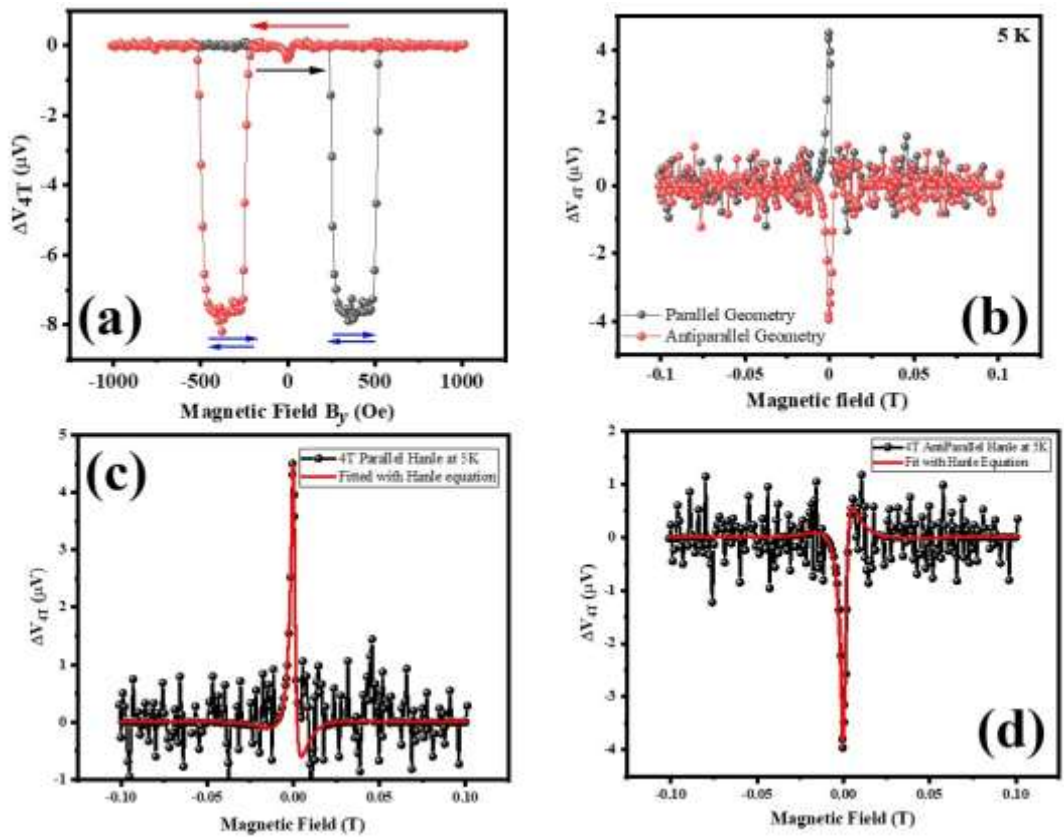

**Figure 3. (a)** Spin valve signal recorded at 5 K in nonlocal device. **(b)** Hanle signal recorded at 5 K in nonlocal 4-T Hanle device for parallel and antiparallel configuration. The solid curves in **(c)** and **(d)** correspond to the best fit to Eq. (4).

We observed a small cusp-like feature around zero magnetic field, as shown in Fig. 4(a), which has already been reported earlier in nonlocal spin devices.[45] The reason has been attributed to enhanced spin dephasing caused by residual magnetostatic fringe fields, hyperfine-field-induced dephasing, or weak localization of spins near the injector interface. Importantly, the cusp is symmetric with respect to magnetic field reversal and does not show hysteresis, ruling out conventional magnetoresistive artifacts such as AMR or local Hall effects.

Figure 3(b) shows the Hanle-type spin precession measured in the nonlocal magnetoresistance (NL-MR) configuration at 5 K, both in parallel and antiparallel configuration. The spin precession signal is recorded by sweeping a magnetic field applied perpendicular to the substrate plane (along the z direction). A clear Hanle line shape with a low noise level is observed, confirming controlled spin precession and dephasing in the silicon channel. The experimental data are well fitted using Eq. (4), (as shown in Fig. 3(c) and (d)), which describes Hanle spin precession under diffusive spin transport. During the measurement, the applied out-of-plane magnetic field remains smaller than the coercive fields of the ferromagnetic electrodes. Consequently, the magnetization directions of both $Mn_2CoAl$ electrodes remain aligned in the plane of the substrate throughout the field sweep. Under this condition, the observed Hanle response originates solely from spin precession in the silicon channel, and the excellent agreement between the experimental data and Eq. (4) validates the applicability of the diffusive spin transport model.

From the fitting of the nonlocal Hanle curve at 5 K, the extracted spin transport parameters are tabulated below.

**Table 1.** Extracted spin transport parameters from the fitting of the nonlocal Hanle curve detected at 5 K.

| Temperature (K) | Device Geometry | Configuration | Spin lifetime (ns) | Spin diffusion length (μm) | Spin polarization |
|---|---|---|---|---|---|
| 5 | Nonlocal Hanle | Parallel | 5.65 | 0.82 | 0.0152 |
|   |   | Antiparallel | 4.98 | 0.78 | 0.0149 |

The observation of a well-defined nonlocal Hanle precession signal, together with the spin-valve effect, provides compelling evidence of electrical spin injection, transport, and detection in the p-Si channel. Long spin lifetimes and extended spin diffusion lengths in p-type silicon are crucial for efficient spin transport and device scalability in semiconductor spintronics. Although the intrinsic spin lifetime is determined by the bulk properties of silicon, the experimentally observed spin lifetime is strongly influenced by the ferromagnetic tunnel contact through interfacial spin-flip scattering and magnetostatic effects. The use of the spin-gapless semiconductor $Mn_2CoAl$ suppresses such extrinsic relaxation channels due to its high spin polarization and improved conductivity matching, enabling more faithful access to the intrinsic spin transport properties of p-Si.

From Fig. 3 (b), we can see that the spin accumulation amplitude in the parallel configuration is slightly larger than that in the antiparallel configuration. This asymmetry can be attributed to the difference in spin-dependent electrochemical potential alignment at the detector contact for the two magnetization states. Additional contributions from slight asymmetry in interface spin polarization, spin-dependent tunneling efficiency, and residual magnetostatic fringe fields[39,46,47] can further enhance this difference. Importantly, the presence of well-defined Hanle precession in both configurations confirms that the observed signals originate from genuine spin transport in the silicon channel rather than from spurious magnetoresistive effects.

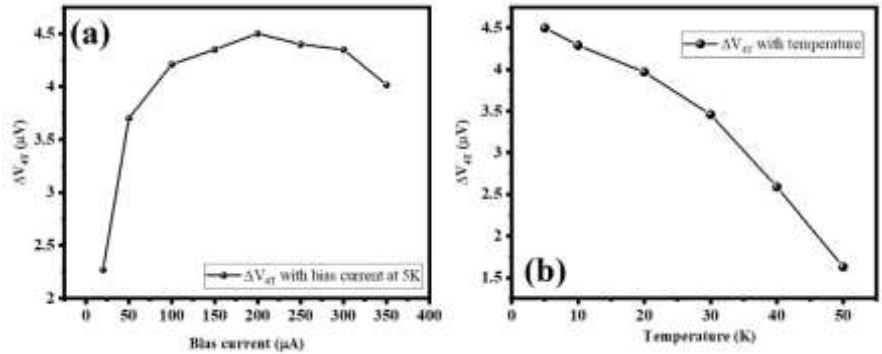

**Figure 4. (a)** Bias current and **(b)** temperature dependence of the nonlocal spin accumulation voltage.

Figure 4(a) presents the bias current dependence of the nonlocal spin accumulation voltage $\Delta V_{4T}$ measured at 5 K. At low injection currents, $\Delta V_{4T}$ increases with bias, reflecting the enhancement of spin accumulation due to increased spin-polarized carrier injection into the p-Si channel. As the injected current is raised, the nonlocal signal reaches a maximum at approximately 200 µA. Beyond this optimal bias, $\Delta V_{4T}$ decreases with further increase in current. This reduction is attributed to bias-induced degradation of the effective spin polarization at the $Mn_2CoAl/SiO_2$/p-Si interface, arising from enhanced spin-flip scattering,[48-50] local Joule heating, and increased occupation of inelastic tunneling channels. These effects reduce the efficiency of spin injection and detection, despite the higher charge current. The observed nonmonotonic bias dependence is characteristic of tunnel-mediated spin injection in semiconductor channels and indicates the presence of an optimal bias window for maximizing nonlocal spin signals.

Figure 4(b) shows the temperature dependence of $\Delta V_{4T}$ extracted from the nonlocal Hanle measurements. With increasing temperature, $\Delta V_{4T}$ decreases monotonically, indicating a progressive reduction of the spin accumulation in the p-Si channel. This behavior is expected, as the nonlocal signal amplitude scales with the product of the spin polarization at the injector and detector interfaces and the spin diffusion length in the semiconductor, all of which are reduced at elevated temperatures. Above approximately 50 K, the nonlocal Hanle signal becomes comparable to the noise level, preventing reliable extraction of spin transport parameters. This suppression is attributed to the combined effects of thermal reduction of interfacial spin polarization at the $Mn_2CoAl/SiO_2$/p-Si junction, enhanced spin relaxation in the semiconductor channel, and the intrinsically small magnitude of nonlocal voltages, which are particularly susceptible to thermal and electrical noise. Similar limitations in observing nonlocal spin signals at elevated temperatures have been widely reported in silicon-based spin transport studies and highlight the critical role of interface quality and thermal stability for achieving robust nonlocal spin detection.

In conclusion, we demonstrate electrical spin injection and long-range spin transport in lightly doped p-type silicon using the spin-gapless Heusler compound $Mn_2CoAl$ as a spin injector. Robust three-terminal Hanle signals persist up to room temperature, yielding a spin lifetime of ~0.68 ns at 300 K that increases to 4.11 ns at 5 K, exceeding typical values reported for conventional ferromagnetic contacts. Temperature-dependent measurements suggest that spin relaxation is governed primarily by hole-mediated and interface-related mechanisms. Moreover, clear nonlocal spin-valve and Hanle signals observed at 5 K provide unambiguous evidence of diffusive spin transport, with an extracted spin lifetime of ~5.65 ns and diffusion length of ~0.82 μm. The improved spin transport is attributed to the high spin polarization and favorable conductivity matching of $Mn_2CoAl$, which helps mitigate interfacial spin depolarization. Although nonlocal signals diminish at elevated temperatures, these results establish spin-gapless semiconductors as effective spin injectors for silicon and offer a promising route toward Si-compatible spintronic devices.

The authors acknowledge the Central Research Facility (CRF), National Institute of Technology Rourkela, for access to experimental characterization facilities. The corresponding author (N.M.) sincerely acknowledges financial support from the Seed Grant program of the National Institute of Technology Rourkela.

## AUTHOR DECLARATIONS

### Conflict of Interest

The authors have no conflicts to disclose.

**Nilay Maji:** Conceptualization (lead); Data curation (lead); Formal analysis (lead); Methodology (lead); Resources (lead); Software (lead); Validation (lead); Visualization (lead); Writing – original draft (lead); Writing – review & editing (equal). Project administration; Project Supervision. **Subham Mohanty:** Formal analysis (equal); Software (equal). **Pujarani Dehuri:** Formal analysis (equal); Software (equal), **Garima Yadav:** Formal analysis (equal); Software (equal).

## DATA AVAILABILITY

The data that support the findings of this study are available from the corresponding author upon reasonable request.